%% file: samplepaper.tex
\documentclass[runningheads]{llncs}
\usepackage{graphicx}
\usepackage[utf8]{inputenc}
\usepackage{amsmath}
\usepackage{graphicx}
\usepackage{float}
\usepackage{subfig}
\usepackage{multirow}
\usepackage{hyperref}
\usepackage{multicol}
\usepackage[colorinlistoftodos]{todonotes}
\usepackage{comment}
\usepackage[boxed,ruled,commentsnumbered]{algorithm2e}
\usepackage{verbatim}
\usepackage{algpseudocode}
\usepackage{ifthen}
\usepackage{multicol}
\usepackage{amsmath,amssymb}
\usepackage{booktabs}
\usepackage{paralist}
\usepackage{multicol}
\usepackage{multirow}
\usepackage{todonotes}
\usepackage{boxedminipage}
\usepackage{times}

\usepackage[para]{footmisc}

\newcommand{\term}[1]{{\small\sf #1}}
\newcommand{\aspect}[1]{\textsf{\itshape #1}}
\newcommand{\triple}[1]{$\langle${\small\tt #1}$\rangle$}

\newcommand{\sbjcomment}[1]{}
\begin{document}
\title{Aspect-based Academic Search using \\Domain-specific KB}
\titlerunning{Aspect-based Academic Search using Domain-specific KB}
% If the paper title is too long for the running head, you can set
% an abbreviated paper title here
%
%\begin{comment}
\author{Prajna Upadhyay \inst{1} \and
Srikanta Bedathur \inst{1} \and
Tanmoy Chakraborty \inst{2} \and Maya Ramanath \inst{1}}
\authorrunning{P. Upadhyay et al.}
% First names are abbreviated in the running head.
% If there are more than two authors, 'et al.' is used.
%
\institute{IIT Delhi, Hauz Khas, New Delhi-110016 \and IIIT-Delhi, Okhla Industrial Estate, Phase III, New Delhi-110020\\
%\email{lncs@springer.com}\\
%\url{http://www.springer.com/gp/computer-science/lncs} \and
%ABC Institute, Rupert-Karls-University Heidelberg, Heidelberg, Germany\\
\email{\inst{1}\{prajna.upadhyay,srikanta,ramanath\}@cse.iitd.ac.in},\\
\inst{2}\email{tanmoy@iiitd.ac.in}}
%tanmoy@iiitd.ac.in \inst{2}}
%\end{comment}
%
\maketitle              % typeset the header of the contribution
\begin{abstract}
Academic search engines allow scientists to explore related work relevant to a given query. Often, the user is also aware of the \emph{aspect} to retrieve a relevant document. In such cases, existing search engines can be used by expanding the query with terms describing that aspect. However, this approach does not guarantee good results since plain keyword matches do not always imply relevance. To address this issue, we define and solve a novel academic search task, called \emph{aspect-based retrieval}, which allows the user to specify the aspect along with the query to retrieve a ranked list of relevant documents. The primary idea is to estimate a language model for the aspect as well as the query using a domain-specific knowledge base and use a mixture of the two to determine the relevance of the article. Our evaluation of the results over the Open Research Corpus dataset shows that our method outperforms keyword-based expansion of query with aspect with and without relevance feedback.
\end{abstract}
%\vspace*{-5mm}
\input{introduction.tex}
\input{model.tex}
\input{experiments.tex}

\input{conclusion.tex}

%\begin{thebibliography}{8}
\bibliographystyle{splncs04}
\small
\bibliography{samplepaper}

\end{document}

%% file: introduction.tex
%\vspace*{-5mm}
\section{Introduction}
%\vspace*{-3mm}
\label{sec:intro}
Academic search engines such as Google Scholar,
%,\footnote{\url{https://scholar.google.com}} 
PubMed,
%\footnote{\url{https://ncbi.nlm.nih.gov/pubmed/}} 
and Semantic Scholar
%,\footnote{\url{https://semanticscholar.org}} 
play a central role in the lives of researchers dealing with the ever growing flood of related work. To further improve the academic search experience, there have been proposals to either re-rank the results using user's interests~\cite{Sugiyama:2015:CES:2785536.2785566} and the set of papers assessed relevant~\cite{doi:10.1108/EL-04-2017-0077}, or to recommend new articles based on a query article~\cite{Chakraborty:2016:FFR:2990640.2990686}.

In this paper, we define and solve a novel academic search task, called \emph{aspect-based retrieval}, which is targeted towards enabling the academic search user to specify the \emph{aspect} along with the query to retrieve a ranked list of scientific articles that are (i) relevant to the query, and (ii) the relevance relation \cite{bean2001} between the query and retrieved documents is semantically close to the specified aspect. We illustrate this expected behavior with a concrete example: consider the query \term{\footnotesize{autoencoder}} and an aspect of interest, say, \aspect{\footnotesize{application}}, then we aim to rank high the articles which are related to the concept of \term{\footnotesize{autoencoder}} \emph{and} are about the \textit{applications} of autoencoders. If there were two papers, titled (a) ``Complex-valued Autoencoders'' and (b) ``Exploring autoencoders for unsupervised feature selection'', our system should rank the paper (b) higher than paper (a) since it specifically deals with applications of the specified query rather than its variants. Note that both papers are indeed relevant to the central concept being queried. Aspect-based retrieval of scientific documents is not straightforward at all since the relation semantics do not manifest as simple keyword matches. Simply expanding the query with terms that define the aspect fails to retrieve relevant articles. In the above example, the paper (b) does not contain the \term{\footnotesize{application}} in the title as well as the abstract. On the other hand, a document, titled ``Evaluating the Performance of Dynamic Database Applications'' is not related to the query \term{\footnotesize{ RDBMS}} along the \aspect{\footnotesize{application}} aspect although its title contains the term \term{\footnotesize{applications}}. The way in which the specified semantic relationship manifests between the query and a document is highly dependent on the domain we operate in.
For example, given the query \term{\footnotesize{autoencoder}}, for a document to be related along \aspect{\footnotesize{application}} aspect, the presence of terms like \term{\footnotesize{feature}} or \term{\footnotesize{selection}} with terms similar to \term{\footnotesize{application}} would be highly suggestive because \term{\footnotesize{feature\_selection}} is known to be an application of \term{autoencoder}. But this can not be easily determined by just analysing the documents without prior knowledge of the domain.
%%%%%%%%%%%%%%%%%%%%

%%%%%%-----------------As we show later, this is not just an idiosyncratic example. The problem appears in many other cases. It should be noted that although our work closely resembles the faceted retrieval~\cite{Gupta:2018:GSA:3197026.3203900,Wu:2011:IAW:2016945.2016963}, unlike faceted retrieval systems it does not generate and evaluate of quality of facets given a query. Instead we focus on ranking relevant documents given a query and a facet. In academic search, techniques have been proposed to retrieve relevant papers along their citation context in a query document~\cite{Chakraborty:2016:FFR:2990640.2990686}.

%\subsection{Our Approach}
We address the challenge of aspect-oriented retrieval for scientific documents by minimizing the risk of returning a document whose language model diverges from the model estimated for domain-specific query and aspect specified by the user. The query and aspect models are derived using domain-specific knowledge bases (KBs), which is challenging in itself due to the inherent sparsity of relations in these KBs. By using a domain-specific KB of computer science, like TeKnowbase \cite{Upadhyay:2018:CAT:3184558.3191532}, we show how to overcome the sparsity issue in KB via inference using meta-paths derived from the KB. Our results over the Open Research Corpus~\cite{ammar:18} dataset containing more than 39 million published research papers show that our proposed approach outperforms variants of query likelihood language models with/without relevance feedback. 

%% file: model.tex
\section{Aspect based retrieval model}
%\vspace*{-2mm}
\label{sec:model}
\begin{comment}
\begin{figure}[!h]
\centering
\subfloat[] {\includegraphics[width=0.2\textwidth]{figs/q_ind.pdf}} \hfill
\subfloat[] {\includegraphics[width=0.3\textwidth]{figs/q_dep.pdf}}\hfill
\subfloat[] {\includegraphics[width=0.35\textwidth]{figs/scoring.pdf}}\hfill
\caption{Components of the mixture distribution. a) The query independent component defines a probability distribution $P(w|a)$ on terms based on the aspect alone. For \aspect{application} aspect, terms like \term{application} or \term{usage} are assigned high probabilities. b) The query and aspect dependent component uses relationships in a domain specific knowledge graph to assign probabilities $P(w|q,a)$ to terms. For a query \term{autoencoder} and \aspect{application} aspect, terms present in entities connected by the \term{application} relation to \term{autoencoder} are assigned high probabilities. c) The final distribution is determined by a mixture of estimated $P(w|a)$ and $P(w|q,a)$.  The documents are ranked in decreasing order of divergence of their language models with that of the estimated mixture model}
\label{fig:system}
\end{figure}
\end{comment}
\subsection{System overview}
%\vspace*{-1mm}
Our system takes a query and an aspect as input, and returns a ranked list of relevant documents. Users express their information need as strings of words called {\bf queries}. In our case, a query is a technical entity. An {\bf aspect} is the relevance relation specified by the user between the query and the relevant document. 
%%%%%%%%%%%
%For example, for the query \term{autoencoder}, a paper titled \emph{Exploring autoencoders for unsupervised feature selection} is more relevant for the \emph{application} aspect.
%%%%%%%
Given a query, the documents are ranked using retrieval models \cite{Manning:2008:IIR:1394399}. A {\bf retrieval model} transforms the document space into an intermediate representation and returns a ranked list of documents according to some scoring function. %Our idea is to estimate a language model \cite{Ponte:1998:LMA:290941.291008} for each aspect and use it as a prior to score the documents.

%Our system takes a query and an aspect as input and returns a ranked list of relevant documents. We use a generative framework to model documents addressing an aspect for a query. Given an aspect, the document is generated according to a probability distribution described by the aspect alone, which is the prior. The query and aspect together model the document according to some other distribution. The final distribution is a mixture of the two. We describe the model and the estimation techniques more formally in the following paragraphs.
%\vspace*{-2mm}
\subsection{Retrieval model and Estimation}
%\vspace*{-2mm}
Language modelling techniques \cite{Ponte:1998:LMA:290941.291008} model the relevance of a document as the probability of generating the query from the document. Expanding the query with aspect terms and doing relevance feedback \cite{Lavrenko:2001:RBL:383952.383972}  will only retrieve documents containing those terms. Given a query $q$ and an aspect $a$, a relevant document $d$ consists of terms determined by both $q$ and $a$. 

\noindent {\bf Aspect dependent prior probability.}
\label{sec:term_prior}
$P(w|a)$ is the prior, which is the probability of a term $w$ appearing in $d$ given an aspect $a$, independent of the query. 

\noindent {\bf Query and aspect dependent probability.}
The probability of a term appearing in $d$ given a query $q$ and aspect $a$ is denoted by $P(w|q,a)$.

\noindent {\bf Mixture of the two probability distributions.}
The relevance of $d$ will be determined by a mixture model of both the probability distributions. Equation \eqref{eq:model_eq} describes our probability distribution. It is denoted as $MM$.
%\vspace*{-2mm}
\small
\begin{equation}
%\vspace*{-2mm}
\label{eq:model_eq}
    MM(w) = \lambda P(w|a) + (1-\lambda)P(w|q,a)
\end{equation}\label{eq2}
\normalsize
\noindent
\noindent {\bf Scoring of documents.}
The language model $M_d$ of a candidate document is expressed by Eq. \eqref{eq3}. Dirichlet smoothing is used for $M_d$.
%\vspace*{-2mm}

\small
 \begin{equation}
 \label{eq3}
M_d(w) = \frac{tf(w,d) + \mu P(w|C)}{length(d)+\mu}
 \end{equation}
 \normalsize
 where $tf(w,d)$ is the frequency of $w$ in $d$, and $P(w|C)$ is the probability of $w$ appearing in the entire collection. The risk associated with using $MM$ to approximate $M_d$ is expressed by KL-divergence between $MM$ and $M_d$ and the documents are returned in an order of increasing KL-divergence.
%\vspace*{-2mm}
\small
\begin{equation}
%\vspace*{-2mm}
\label{eq:kl_divergence}
    KL(MM||M_d) = \sum_{w} P(w|MM)\log\frac{P(w|MM)}{P(w|M_d)}
\end{equation}
\normalsize

%\subsection{Estimation}
%\label{sec:estimation}
%We indexed the Open Research Corpus dataset on Galago to estimate aspect-dependent probability. We later conduct our experiments on the same dataset (described in Section \ref{sec:experiment}).
%The dataset we use....

\begin{figure}[!h]
\vspace*{-3mm}
\centering
\subfloat[] {\includegraphics[width=0.5\textwidth]{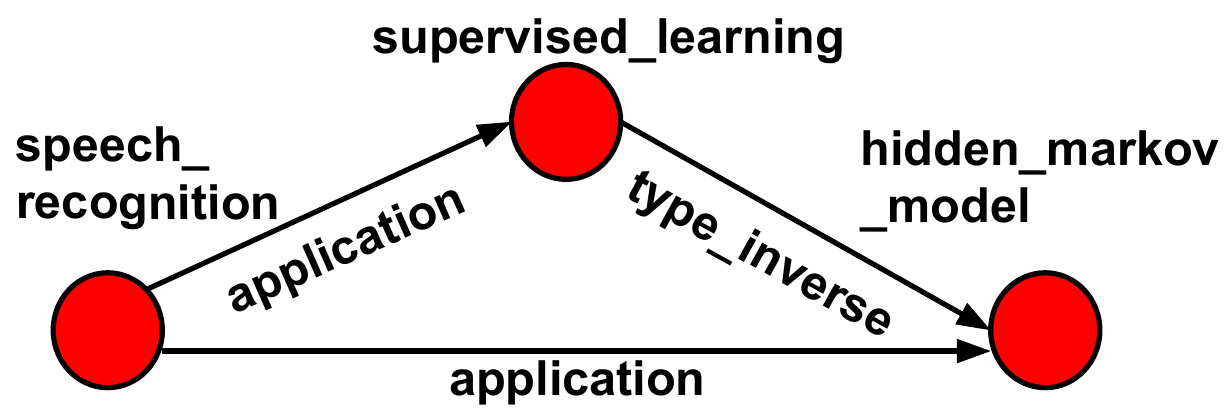}} \hfill
%\subfloat[] {\includegraphics[height=0.9in]{figs/hmm.pdf}} \hfill
\subfloat[] {\includegraphics[width=0.5\textwidth]{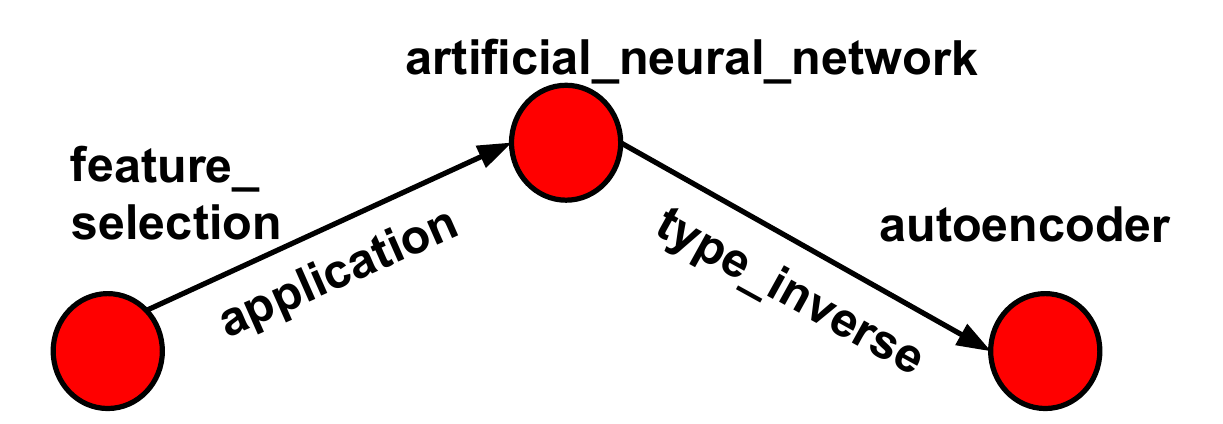}}\hfill
%\subfloat[] {\includegraphics[height=0.9in]{figs/autoencoder_infer.pdf}}\hfill
\caption{Examples of meta-paths for \term{application} relation. (a) The meta-path \triple{application, type\_inverse} exists between \term{speech\_recognition} and \term{hidden\_markov\_model}. (b) \term{feature\_selection} and \term{autoencoder} are not related by \term{application} relation but still it can be inferred because of the existence of the same meta-path \triple{application, type\_inverse} between them.}
\label{fig:meta_path_application}
\vspace*{-4mm}
\end{figure}

\noindent{\bf Estimation of query-independent component.}
\label{sec:query_independent}
$P(w|a)$ is estimated using a narrow set of documents from our dataset acquired as follows. We chose 10 queries and retrieved the top 10 documents for them using the standard query likelihood model. Additionally, we fired queries of the form ``query+aspect" to retrieve the top-10 documents using the same model.  We recruited evaluators to annotate about 1500 documents with the aspect labels (described in details in Section \ref{sec:eval_scheme}). In order to increase the size of our document set, we used heuristics to collect more documents given an aspect. We formulated a query containing only the aspect as a keyword and retrieved documents for it using the standard likelihood model. But, we retained only those documents which contained the name of the aspect in the title on the intuition that such documents are highly likely (though not guaranteed) to be about those aspects. Having a set of ground-truth documents $D$, $P(w|a)$ is estimated according to Equation \eqref{eq:term_prior}.
\small
\begin{equation}
%\vspace*{-2mm}
\label{eq:term_prior}
    P(w|a) = \frac{1}{|D|} \sum_{d \in D} \frac{tf(w,d)}{\sum_{w' \in d}tf(w',d)}
\end{equation}
%\vspace*{-2mm}
\normalsize
%\vspace*{-1mm}
\noindent{\bf Estimation of query-dependent component.}
We used relationships in TeKnowbase ($TKB$)  \cite{Upadhyay:2018:CAT:3184558.3191532} to represent aspects. $TKB$ consists of entities such as \term{\footnotesize{hidden\_markov\_model}} or \term{\footnotesize{ speech\_r\-ecognition}} and other domain-specific relationships like \term{\footnotesize{application}}, \term{implemen\-tation} or \term{algorithm}. 
The triple \triple{speech\_\-recognition, application, hidden\_mar\-kov\_model} conveys information that \term{speech\_recognition} is an application of \term{hidden\_m\-arkov\_model}. The entities connected via \term{application} relation in $TKB$ have a higher probability of appearing in documents addressing \aspect{application} aspect. However, $TKB$ is sparse. To automatically infer the entities participating in a particular relationship type, we used meta-paths. A meta-path is a sequence of edges with labels connecting two nodes which have been used previously for KB completion tasks \cite{Lao:2011:RWI:2145432.2145494}, link prediction \cite{journals/pvldb/Ley09} as well as to find similarity between two nodes \cite{5992571,Shi:2012:RSH:2247596.2247618}. Figure \ref{fig:meta_path_application} shows how meta-paths can be used to infer relationships between entities. 
%\vspace*{-1mm}
%\vspace*{-1mm}

\noindent {\bf 1) Direct inference using meta-paths.}
To automatically determine entities that participate in a given relationship type with entity $e_i$, we used the path-constrained random walk algorithm ($PRA$) proposed in \cite{Lao:2011:RWI:2145432.2145494}. Given a set $E$ of entities in $TKB$, a source node $e_i$ and a meta-path $P$, a path-constrained random walk defines a probability distribution $h_{e_i,P}(e_j)$ to all entities in $E$ which is the probability of reaching $e_j$ from $e_i$ by doing a random walk along $P$. The key idea is to acquire the set of meta-paths representing the given relationship type and use $PRA$ for inferencing. To do so, we retrieved the set of all meta-paths ($MP$) connecting the given relationship type in $TKB$ and scored them according to their frequency. Given a set of meta-paths $MP = P_1, P_2,..., P_n$, the score for each node reachable from source $e_i$ is given by:
$\label{eq:path_constrained}
  score_{e_i}(e_j) =  \alpha_1h_{e_i,P_1}(e_j) + \alpha_2h_{e_i,P_2}(e_j) + ... + \alpha_nh_{e_i,P_n}(e_j),
$
where $\alpha_l (l=1 \ldots n)$ is the frequency of meta-path $P_l$. $score_{e_i}(e_j)$ is converted to a probability distribution using \emph{softmax} and denoted by Equation \eqref{eq:softmax_direct}. \\%Since we have to estimate a probability distribution, we converted $score_{e_i}(e_j)$ to a probability distribution using softmax and denote it as $DI_{e_i}(e_j)$.
%\vspace*{-2mm}

\noindent {\bf 2) Indirect inference using meta-paths.}
\label{sec:metapath2vec}
$PRA$ assigns zero probability to nodes that are not reachable via any meta-paths in $MP$. To address this issue, we used MetaPath2Vec \cite{Dong:2017:MSR:3097983.3098036}. It takes a meta-path as input and constructs embeddings of entities such that entities that are likely to be connected via the meta-path (and not necessarily having a meta-path between them) are assigned vector representations closer to each other.  We used the top-k meta-paths in $MP$ as input to metapath2vec and obtained vector representations $V_e$ for entity $e$. We used the softmax function to convert cosine similarities between entities into a probability distribution as described in Equation \eqref{eq:softmax_indirect}.

\noindent\begin{minipage}{5.7cm}
\small
\begin{equation}
%\vspace*{-2mm}
\label{eq:softmax_direct}
    DI_{e_i}(e_j) = \frac{e^{score_{e_i}(e_j)}}{\sum_{e_{k} \in |E|} e^{score_{e_i}(e_{k})}}
\end{equation}
\end{minipage}
\begin{minipage}{6.5cm}
\small
\begin{equation}
\label{eq:softmax_indirect}
h'_{e_i}(e_j) = \frac{e^{\mathrm{sim}\left(V\left(e_i\right), V\left(e_j\right)\right)}}{\sum_{e_{k} \in |E|} e^{\mathrm{sim}\left(V\left(e_i\right), V\left(e_k\right)\right)}}.
\end{equation}
\end{minipage}
\normalsize

The probability distribution for inferencing is a mixture of $DI_{e_i}(e_j)$ and $h'_{e_i}(e_j)$ using $\beta$ as given in Eq. \eqref{eq:mix_ent_prob}. Since the documents are represented as bag of words, we defined the distribution over terms instead of entities using Eq. \eqref{eq:query_aspect_prob}. $terms(e)$ is the set of words present in the entity $e$, and $e_q$ is the entity that $q$ represents. The final probability is a mixture given by Equation \eqref{eq:model_eq}.
%\vspace{-2mm}
\small
\begin{equation}
\vspace*{-2mm}
\label{eq:mix_ent_prob}
    P_{a}(e_i|e_j) = \beta*DI_{e_i}(e_j) + (1 - \beta)*h'_{e_i}(e_j)
\end{equation}
%\vspace{-2mm}
\begin{equation}
\vspace*{-2mm}
\label{eq:query_aspect_prob}
    P(w|q,a) = \sum_{e}P_{a}(e|e_q), \text{s.t. $w \in terms(e)$}
\end{equation}
%\vspace{-2mm}

\normalsize
%Equation \eqref{eq:term_prior} estimates the prior probability of the term given only the aspect. Equation \eqref{eq:query_aspect_prob} estimates the probability of a term given the query and the aspect both. 

%of $P(w|a)$ and $P(w|q,a)$. $\lambda$ is used to mix the two distributions. The final distribution, as described at the beginning of the section is given below:

%% file: experiments.tex
%\vspace*{-6mm}
\section{Experiments}
\label{sec:experiment}
%\vspace*{-4mm}
\subsection{Setup}
{\bf Dataset.}
We used the Open Research Corpus dataset and indexed it using Galago. The baseline models (described below) are already implemented in Galago.\\
{\bf Aspects.}
We experimented with 3 different aspects -- \aspect{ application}, \aspect{algorithm} and \aspect{implementation}. We set $\lambda$ and $\beta$ to 0.5 in Eqs. \eqref{eq:model_eq} and \eqref{eq:mix_ent_prob}. We restricted ourselves to meta-paths of size at most 3.  We set $k$=5 for choosing the top-k meta-paths for generating embeddings using MetaPath2Vec (described in Section \ref{sec:metapath2vec}).\\
{\bf Benchmarks.}
Benchmark queries were taken from a set of 100 queries released by \cite{Xiong:2017:ESR:3038912.3052558} out of which 43 existed as whole entities in $TKB$, shown in Figure \ref{tab:benchmark}.
\small
\begin{figure*}[]
\vspace*{-5mm}
\centering
\begin{boxedminipage}[c]{\textwidth}
\term{artificial intelligence}, \term{augmented reality},  \term{autoencoder}, \term{big data}, \term{category theory}, 
\term{clojure}, \term{cnn}, \term{computer vision}, \term{cryptography}, \term{data mining},    
\term{data science}, \term{deep learning}, \term{differential evolution}, \term{dirichlet process}, \term{duality},
\term{genetic algorithm}, \term{graph drawing}, \term{graph theory}, \term{hashing}, \term{information geometry},
\term{information retrieval}, \term{information theory}, \term{knowledge graph}, \term{machine learning}, \term{memory hierarchy},
\term{mobile payment}, \term{natural language}, \term{neural network}, \term{ontology}, \term{personality trait},
\term{prolog}, \term{question answering}, \term{recommender system}, \term{reinforcement learning}, \term{sap},
\term{semantic web}, \term{sentiment analysis}, \term{smart thermostat}, \term{social media}, \term{speech recognition},
\term{supervised learning}, \term{variable neighborhood search}, \term{word embedding}
\end{boxedminipage}
\caption{Benchmark queries}
\label{tab:benchmark}
\vspace{-5mm}
\end{figure*}
\vspace{-2mm}
\normalsize

\noindent\textbf{Baselines.}
\label{sec:baseline}
We explicitly added the keyword representing the aspect to the query and used standard retrieval models with/without relevance feedback techniques as baselines described below:\\
{\bf 1) Query likelihood model with query only (QL+query).} Query likelihood \cite{Ponte:1998:LMA:290941.291008} estimates a language model for each document in the collection and ranks them by the likelihood of seeing the query terms as a random sample given that document model. 
\begin{comment}
Following is a brief overview of query-likelihood model. Using Bayes rule, we can write he likelihood of a document given a query as follows:

\begin{equation*}
    P(d|q) = \frac{P(q|d)P(d)}{P(q)}
\end{equation*}
Since $P(q)$ is the same for all the documents, we can ignore this term. $P(d)$ can be assumed to be uniform for all the documents. So, this can be ignored also. The final likelihood is given by Equation \eqref{eq:query_likelihood}. It is dirichlet smoothed with parameter $\mu$.

\begin{equation*}
    P(d|q) = P(q|d)
\end{equation*}

$P(q|d)$ is estimated as follows:

%\vspace*{-2mm}
\small
\begin{equation}
\label{eq:query_likelihood}
    P(d|q) \propto P(q|d) = \prod_{w \in terms(q)}\frac{tf(w)+\mu P(w|C)}{|d|+\mu}
\end{equation}
\normalsize
\end{comment}
\\
 {\bf 2) Query likelihood model with query + aspect name (QL + query + aspect).} We used the same model as above but added the terms \term{application}, \term{algorithm} or \term{implementation} to the query based on the aspect and retrieved the results. \\
 {\bf 3) Query expansion with pseudo relevance feedback on  QL + query + aspect (QL + query + aspect + QE).} We chose top-100 terms to expand the query for the query used in the previous baseline using relevance feedback model \cite{Lavrenko:2001:RBL:383952.383972}. Top-1000 documents were used as feedback documents. The weight of the original query was set as 0.75.
 
\noindent{\bf 4) Mixture Model (MM)}. This is our retrieval model described in Section \ref{sec:model}.
\subsection{Evaluation scheme and metrics}
%\vspace*{-5mm}
{\bf Evaluation scheme.} 
\label{sec:eval_scheme}
In the absence of an extensive ground-truth dataset, we conducted a crowd-sourced user-evaluation exercise (involving Computer Science students and researchers, not related to the project) to measure the performance of our model. %We built a web-based interface to facilitate the evaluations. We recruited students and researchers of Computer Science (not directly related to the project) to conduct the evaluations. They were asked to evaluate the top-5 documents returned by our method as well as competing techniques for pairs of a given query and an aspect. Because determining if a paper addresses a particular aspect is a subjective question, we formulated domain specific rules with the help of in-house experts to determine the relationship between the query and the retrieved paper. 
We formulated domain-specific questions, and depending on the answers marked by the evaluators, the documents were assigned a particular score for a query and aspect pair.
\\
\noindent{\bf Evaluation metrics.}
Each query and abstract pair was graded by at least 2 evaluators. We converted the response from each of them into a graded relevance scale and averaged the relevance values marked by them for each query-abstract pair. We used {\bf Discounted Cumulative Gain (DCG)} and {\bf Precision} to evaluate top-5 documents.
%\vspace*{-4mm}
\subsection{Results and Discussions}
%\vspace*{-3mm}
Table \ref{tab:algorithm_results} shows the results for \aspect{algorithm}, \aspect{application} and \aspect{implementation} aspects. We observe that our model outperforms the rest of the baselines in terms of precision@5 and DCG for all of the 3 aspects. {\bf QL + query + aspect + QE} comes second after our retrieval model. By modelling the aspect and query dependant probability explicitly, we were able to address the problems of simple keyword-based match for aspects described in Section \ref{sec:intro} and \ref{sec:model}. For example, the top-2 papers retrieved for \term{genetic\_algorithm} for application aspect by our model were \emph{Genetic Ant Algorithm for Continuous Function Optimization and Its MATLAB Implementation} and \emph{Solve Zero-One Knapsack Problem by Greedy Genetic Algorithm}. The top-2 papers retrieved by {\bf QL + query + aspect + QE} for \aspect{application} aspect do not describe any application of \term{genetic\_algorithm} but contained a few terms like ``a wide application prospect" in the abstract due to which it was retrieved in the top positions. Adding relevance feedback terms also did not work because the list of pseudo relevant documents did not contain relevant documents in the first place due to plain keyword-based retrieval. Both the papers retrieved by our method address application aspect for \term{genetic\_algorithm} even if ``application" is not mentioned in the title. 
\vspace{-4mm}
\begin{table*}[t]
\caption{Results for algorithm, application and implementation aspect.}
\label{tab:algorithm_results}
\centering
\resizebox{\textwidth}{!}{\begin{tabular}{p{3cm}p{1.25cm}p{0.75cm}p{0.75cm}p{1.25cm}p{0.75cm}p{0.75cm}p{1.25cm}p{0.75cm}p{0.75cm}}
\toprule
%{\bf & & & & & & \\ \hline}
& \multicolumn{3}{c}{\textbf{Algorithm}} & \multicolumn{3}{c}{\textbf{Application}} & \multicolumn{3}{c}{\textbf{Implementation}} \\
\cmidrule(lr){2-4}\cmidrule(lr){5-7}\cmidrule(lr){8-10}
{\bf Approach} &  {DCG@5} &  {P@5}  & {P@1} &  {DCG@5} &  {P@5}  & {P@1} &  {DCG@5} &  {P@5} & {P@1}  \\ \midrule
{\bf MM} & {\bf 6.27} & {\bf 0.70} & {\bf 0.75}  & {\bf 2.64} & {\bf 0.45} & {\bf 0.47} & {\bf 2.33} & {\bf 0.44} & 0.40 \\ 
{\bf QL+query} & 2.69 & 0.3 & 0.33 & 1.42 & 0.25 & 0.22 & 1.05 & 0.16 & 0.23 \\ 
{\bf QL+query+aspect} & 5.03 & 0.56 & 0.59 & 2.38 & 0.41 & 0.35 & 1.92 & 0.30 & \em{0.43} \\ 
{\bf QL+query+aspect+QE} & \em{5.12} & \em{0.58} & \em{0.61} &  \em{2.5} & \em{0.43} & \em{0.41} & \em{2.29} & \em{0.37} & {\bf 0.49} \\ \bottomrule
\end{tabular}}

%\vspace{-5mm}
\end{table*}
\normalsize

%% file: conclusion.tex
%\vspace*{-4mm}
\section{Conclusion}
\label{sec:con}
%\vspace{-2mm}
In this paper, we built an aspect-based retrieval model for scientific literature using TeKnowbase. Given a query and an aspect, this model returns a ranked list of documents that address that aspect for the query. We tested our model for 43 queries and 3 aspects with satisfactory results. We could beat the results obtained by adding aspect name explicitly to the query and doing pseudo-relevance feedback on those documents. 

\noindent\textbf{Acknowledgements}\\
This work was partially supported by IIT Delhi-IBM Research AI Horizons Network collaborative grant;  Ramanujan Fellowship, DST (ECR/2017/00l691) and the Infosys Centre for AI, IIIT-Delhi, India. T. Chakraborty would like to thank the support of Google India Faculty Award.

%% file: samplepaper.bbl
\begin{thebibliography}{10}
\providecommand{\url}[1]{\texttt{#1}}
\providecommand{\urlprefix}{URL }
\providecommand{\doi}[1]{https://doi.org/#1}

\bibitem{ammar:18}
Ammar, W., Groeneveld, D., Bhagavatula, C., Beltagy, I., Crawford, M., Downey,
  D., Dunkelberger, J., Elgohary, A., Feldman, S., Ha, V., Kinney, R.M.,
  Kohlmeier, S., Lo, K., Murray, T.C., Ooi, H.H., Peters, M.E., Power, J.L.,
  Skjonsberg, S., Wang, L.L., Wilhelm, C., Yuan, Z., van Zuylen, M., Etzioni,
  O.: {Construction of the Literature Graph in Semantic Scholar}. NAACL  (2018)

\bibitem{doi:10.1108/EL-04-2017-0077}
Aravind Sesagiri~Raamkumar, S.F., Pang, N.: {Can I have more of these please?:
  Assisting researchers in finding similar research papers from a seed basket
  of papers}. The Electronic Library  (2018)

\bibitem{bean2001}
Bean, C., Green, R.: {Relevance Relationships}. Information Science and
  Knowledge Management  (2001)

\bibitem{Chakraborty:2016:FFR:2990640.2990686}
Chakraborty, T., Krishna, A., Singh, M., Ganguly, N., Goyal, P., Mukherjee, A.:
  {FeRoSA: A Faceted Recommendation System for Scientific Articles}. PAKDD
  (2016)

\bibitem{Dong:2017:MSR:3097983.3098036}
Dong, Y., Chawla, N.V., Swami, A.: {Metapath2Vec: Scalable Representation
  Learning for Heterogeneous Networks}. KDD  (2017)

\bibitem{Lao:2011:RWI:2145432.2145494}
Lao, N., Mitchell, T., Cohen, W.W.: {Random Walk Inference and Learning in a
  Large Scale Knowledge Base}. EMNLP  (2011)

\bibitem{Lavrenko:2001:RBL:383952.383972}
Lavrenko, V., Croft, W.B.: {Relevance Based Language Models}. SIGIR  (2001)

\bibitem{journals/pvldb/Ley09}
Ley, M.: {DBLP - Some Lessons Learned.} PVLDB  (2009)

\bibitem{Manning:2008:IIR:1394399}
Manning, C.D., Raghavan, P., Schütze, H.: {Introduction to Information
  Retrieval}. Cambridge University Press (2008)

\bibitem{Ponte:1998:LMA:290941.291008}
Ponte, J.M., Croft, W.B.: {A Language Modeling Approach to Information
  Retrieval}. SIGIR  (1998)

\bibitem{Shi:2012:RSH:2247596.2247618}
Shi, C., Kong, X., Yu, P.S., Xie, S., Wu, B.: {Relevance Search in
  Heterogeneous Networks}. EDBT  (2012)

\bibitem{Sugiyama:2015:CES:2785536.2785566}
Sugiyama, K., Kan, M.Y.: A comprehensive evaluation of scholarly paper
  recommendation using potential citation papers. Int. J. Digit. Libr.  (2015)

\bibitem{5992571}
Sun, Y., Barber, R., Gupta, M., Aggarwal, C.C., Han, J.: {Co-author
  Relationship Prediction in Heterogeneous Bibliographic Networks}. ASONAM
  (2011)

\bibitem{Upadhyay:2018:CAT:3184558.3191532}
Upadhyay, P., Bindal, A., Kumar, M., Ramanath, M.: Construction and
  applications of teknowbase: A knowledge base of computer science concepts.
  Companion Proceedings of the The Web Conference  (2018)

\bibitem{Xiong:2017:ESR:3038912.3052558}
Xiong, C., Power, R., Callan, J.: {Explicit Semantic Ranking for Academic
  Search via Knowledge Graph Embedding}. WWW  (2017)

\end{thebibliography}
